\documentclass[a4paper]{jpconf}
\usepackage{graphicx}
\usepackage{lineno}
\usepackage{hyperref}
\usepackage{subfigure}

\begin{document}
%\linenumbers

\title{Explorations of the viability of ARM and Xeon Phi for physics processing}

\author{David Abdurachmanov$^1$, Kapil Arya$^2$, Josh Bendavid$^3$, Tommaso Boccali$^4$, Gene Cooperman$^2$, Andrea Dotti$^5$, Peter Elmer$^6$, Giulio Eulisse$^7$, Francesco Giacomini$^8$, Christopher D. Jones$^7$, Matteo Manzali$^8$, Shahzad Muzaffar$^7$}

\address{$^1$ Digital Science and Computing Center, Faculty of Mathematics and Informatics, Vilnius University, Vilnius, Lithuania}
\address{$^2$ College of Computer and Information Science, Northeastern University, Boston, MA, USA}
\address{$^3$ European Organisation for Nuclear Research (CERN), Geneva, Switzerland}
\address{$^4$ INFN Sezione di Pisa, Pisa, Italy}
\address{$^5$ SLAC National Accelerator Laboratory, Menlo Park, CA 94025, USA}
\address{$^6$ Department of Physics, Princeton University, Princeton, NJ 08540, USA}
\address{$^7$ Fermilab, Batavia, IL 60510, USA}
\address{$^8$ INFN-CNAF, Bologna, Italy}
%\address{$^8$ INFN-CNAF, Viale Berti Pichat 6/2, Bologna 1-40127, Italy}
\ead{Peter.Elmer@cern.ch}

\begin{abstract}
We report on our investigations into the
viability of the ARM processor and the Intel Xeon Phi co-processor for
scientific computing. We describe our experience porting
software to these processors and running benchmarks using real physics
applications to explore the potential of these processors for
production physics processing. 
\end{abstract}

\section{Introduction}

The computing requirements for high energy physics (HEP) projects like the 
Large Hadron Collider (LHC)~\cite{LHCMACHINE} at the European
Laboratory for Particle Physics (CERN) in Geneva, Switzerland are
larger than can be met with resources deployed in a single computing
center. This has led to the construction of a global distributed computing 
system known as the Worldwide LHC Computing Grid (WLCG)~\cite{WLCG},
%, which as
%of 2013 provides $\sim$350k x86-cores and 200PB of disk storage to the
%LHC experiments to run more than 2 million jobs per day. 
which brings together resources from nearly 160 computer centers in 35 
countries. Computing at this scale has been used, for example, 
by the CMS~\cite{CMSDET} and ATLAS~\cite{ATLASDET} experiments for the 
discovery of the Higgs boson~\cite{CMSHIGGS,ATLASHIGGS}. To achieve this
and other results the CMS experiment, for example, typically used during 
2012 a processing capacity between 80,000 and 100,000 x86-64 cores from
the WLCG.
Further discoveries are possible in the next decade as the LHC moves to 
its design energy and increases the machine luminosity. However, increases in
dataset sizes by 2-3 orders of magnitude (and commensurate processing
capacity) will eventually be required to realize the full potential of this
scientific instrument. The scale and longevity of the LHC computing 
require continual R\&D into new technologies which may be relevant in
the coming years. 
Since around 2005 processors have also hit scaling limits, largely
driven by overall power consumption~\cite{GAMEOVER}, which have led
to the introduction of multicore CPUs and which are driving interest
in processor architectures other than simple, general purpose x86-64
processors. In this paper we report on our investigations into two
such technologies: low power ARM processors~\cite{ARMPROC} and Intel's
Xeon Phi coprocessor card, for scientific computing. For ARM, we extend and
build on our previously reported results~\cite{ACAT13ARM}.

\section{ARM Investigations}
\subsection{Test Setup}

For the tests described in this paper we used two low-cost development boards,
the ODROID-U2 (purchased in Feb. 2013) and the ODROID-XU+E (purchased in
Aug. 2013)~\cite{ODROID}. The processor on the U2 board is an 
Exynos 4412 Prime, a System-on-Chip (SoC) produced by Samsung for use
in mobile devices. It is a quad-core Cortex A9 ARMv7 processor operating
at 1.7GHz with 2GB of LP-DDR2 memory. The processor also contains an ARM 
Mali-400 quad-core GPU accelerator, although that was not used for the work
described in this paper. The XU+E board has a more recent Exynos 5410
processor, with 4 Cortex-A15 cores at 1.6GHz and 4 Cortex-A7 cores at 1.2GHz,
in ARM's big.LITTLE configuration, with 2GB of LDDR3 memory, as well as
a PowerVR SGX544MP3 GPU (also not used in this work).
Both boards have eMMC and microSD slots, multiple
USB ports and 10/100Mbps Ethernet with an RJ-45 port. Power is provided via
a 5V DC power adaptor. ARM's big.LITTLE heterogeneous architecture
in principle pairs each performance (A15) core with a low-power (A7)
core to facilitate a more flexible performance/power response
than simple dynamic clock frequency scaling. The architecture allows for
a mix of performance and low-power cores, however the implementation
in the current generation 5410 chip only allows for switching
between the entire A15 core cluster and the entire A7 cluster. The
other unique aspect of the XU+E board is integrated power sensors
providing access to the power individually used by the A15 cores,
the A7 cores, the GPU and the memory.
The cost of the U2 board alone was \$89 and with the relevant
accessories (cables, a cooling fan, a 64GB eMMC storage module, etc.)
the total cost was \$233. The XU+E board alone cost \$199 and
with the adaptor, 64GB eMMC storage and a ``Smart Power'' meter the
total cost was \$357. The ``Smart Power'' meter allows external
measurements of the total power use of either board, with the values
being made available via USB to the board itself.
The extremely modest cost of these boards permitted us to do
meaningful initial investigations without investing in full-fledged servers.

For the Linux operating system on the U2 (XU+E) board we used
Fedora 18 (19) ARM Remix
with kernel version 3.0.75 (3.4.5), provided by Hardkernel,
the vendor for the ODROID boards. We chose Fedora due to its similarities 
to Scientific Linux CERN (SLC). It is fully hard float
capable and uses the floating point unit on the SoC. The kernel was
reconfigured to enable swap devices/files, which is required for
CMSSW compilation.
%A 4GB swap file was used in our build environment.
All build tests were done using a 500GB $3.5\,''$ ATA disk connected
via USB or with the eMMC storage. Run-time tests were done with output
written to the eMMC storage.

In order to compare results from the ARM board we also used two typical
x86-64 servers currently deployed at CERN. The first is a dual quad-core
Intel Xeon L5520 $@$ 2.27 GHz (Nehalem) with 24GB of memory. The second
is dual hexa-core Intel Xeon E5-2630 $@$ 2.00GHz (Sandy Bridge) with 64GB
of memory. Both machines were equipped with a large local disk for
output and used software installed on an AFS filesystem at CERN.
These machines were
purchased about three years apart and very roughly represent the range
of x86-64 hardware being operated at the time of our ARM tests.

\subsection{Software Environment}
The software environment used here was as described in our earlier
results~\cite{ACAT13ARM}, except that we updated to a newer CMS
software version: a nearly-final pre-release of CMSSW\_7\_0\_0.
The main software problem reported earlier was a problem with
ROOT dictionaries. Some bugs and problems have been fixed, but as of
this paper output via ROOT on ARMv7 still does not work. Thus event
output was turned off when running the CMS test application.

\subsection{Experimental Results}
\begin{figure}[tbp]
\centering
\includegraphics[width=0.5\textwidth]{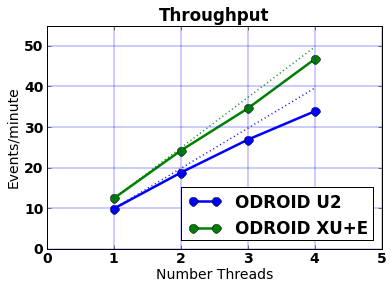}
\caption{Throughput of the multi-threaded ``FullCMS'' application with
varying numbers of threads. Points and solid lines indicate measurements,
dotted lines indicate the extrapolation of the single core results to
show more clearly deviations from scaling.}
\label{fig:arm2}
\end{figure}

We have run again, with the newer CMSSW version, the same test done
previously: a Monte Carlo simulation of 8 TeV LHC minimum bias
events using Pythia8~\cite{PYTHIA8} (event generation) followed by
simulation with Geant4~\cite{GEANT4}. This time we tested both the U2 board
and the XU+E board. This test ran on a single core and we scaled 
the results by the number of physical cores to estimate the
total possible throughput for the chips. In order to load
all cores with a realistic benchmark we ran in addition a second
application using
a beta version of Geant4 version 10 which provides support for event-based multi-threaded applications. We did not use the full CMS simulation for this, but instead a simpler benchmark application (FullCMS) which uses the actual CMS geometry imported from a GDML file. The results are reported in
table~\ref{tab:results}. Here TDP (Thermal Design Power) numbers for the Intel processors were
taken from their website~\cite{XEONTDP} and those for the ODROID
boards were estimated from our own measurements. As before, while
the individual ARM cores are less performant than the x86-64 cores,
a significantly better performance per Watt is obtained. In
figure~\ref{fig:arm2} we show also the performance of the
multi-threaded FullCMS
application with varying numbers of threads. Some deviations from 
linear scaling are seen, especially for the U2 board.

\begin{table}[ht]
\caption{Results of run time tests for single core CMSSW (GEM-SIM)
and a multi-threaded version of the Geant4 benchmark ``FullCMS'' with
4 threads (G4MT).}
\centering
\label{tab:results}
\begin{tabular}{|l|c|c|c|c|c|c|}
  \hline \hline
      &       &             & GEN-SIM & GEN-SIM & G4MT & G4MT \\
      &       &             & Events & Events & Events & Events\\
      &       &    Power    & /minute& /minute & /minute & /minute\\
 Type  & Cores &    (TDP)   & /core  & /Watt  & (threads) & /Watt\\ \hline \hline
 ODROID U2    & $4$ & 4W & 1.08 & 1.08 & 34.2 (4) & 8.6 \\ \hline
 ODROID XU+E  & $4/4$ & 5W & 1.47 & 1.07 & 47 (4) & 9.4 \\ \hline
 dual Xeon L5520     &     &     &     &     &  & \\
 @2.27GHz       & $2\times4$ & 120W & 3.37 & 0.22 & 307.2 (16) & 2.6 \\ \hline
 dual Xeon E5-2630L      &     &     &      &      & & \\
@2.0GHz  & $2\times6$ & 120W & 3.46 & 0.35 & N/A & N/A \\ \hline
\end{tabular}
\end{table}

\subsection{Power investigations}
\begin{figure}[tbp]
\centering
\includegraphics[width=0.75\textwidth]{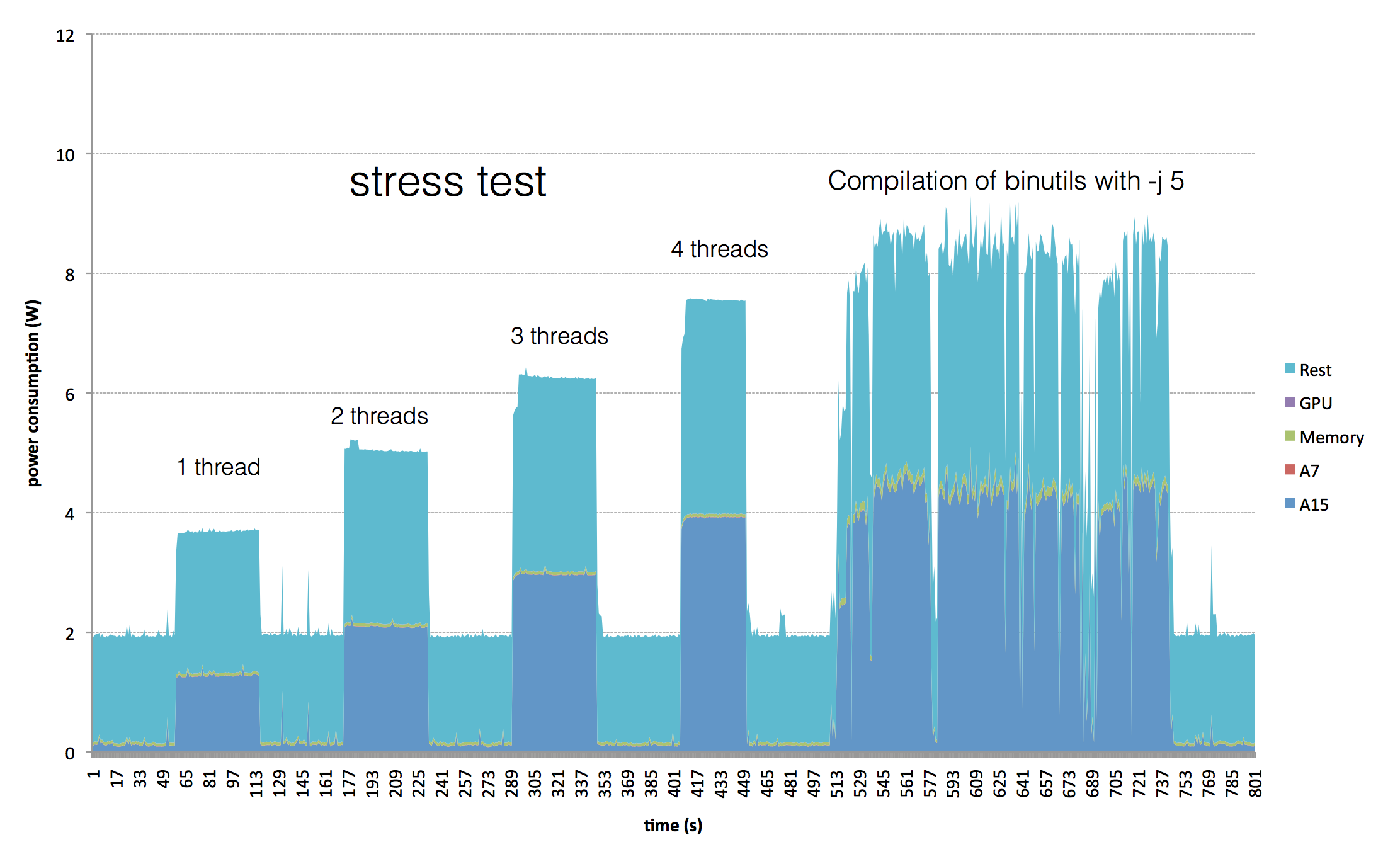}
\caption{Power consumption on the XU+E board as measured by the integrated
power sensors and the external Smart Power meter (labeled ``Rest''), for
tests (from left to right) with {\it stress} using 1,2,3,4 threads and
when compiling a software package (binutils) with ``-j 5''.}
\label{fig:arm1}
\end{figure}

We also performed more detailed investigations into the power
use on the ODROID U2 and XU+E using the ``Smart Power'' meter and
the sensors on the XU+E board. Table~\ref{tab:u2power} shows results
from the ``Smart Power'' meter for the U2 board for various running
conditions. These are consistent with our estimate of $\sim$4W in
table~\ref{tab:results} for the TDP-equivalent power use. In
figure~\ref{fig:arm1} we show power measurements for the XU+E
board when loading 1-4 cores and when compiling. The latter is
consistent with our $\sim$5W TDP equivalent for the XU+E.

\begin{table}[ht]
\caption{Power measurements on the ODROID U2 board using the ``Smart Power''
meter. Here ``fan'' and ``ethernet'' indicates whether the cooling fan and
ethernet were on or not.}
\centering
\label{tab:u2power}
\begin{tabular}{|l|c|c|c|}
  \hline \hline
ODROID U2 & Voltage (V) & Current (mA) & Power (W) \\ \hline \hline
idle, no fan, no ethernet  & 5.02   & 280  & 1.4 \\ \hline

idle, fan, no ethernet  & 5.02   & 360    & 1.8 \\ \hline

idle, no fan, ethernet  & 5.02   & 322  & 1.6 \\ \hline

idle, fan, ethernet  & 5.02   & 400  & 2.0 \\ \hline

full CPU load, no fan, ethernet  & 5.02   & 900  & 4.5 \\ \hline

full CPU load, fan, ethernet  & 5.02   & 970  & 4.9 \\ \hline

\end{tabular}
\end{table}

\section{Xeon Phi Investigations}
As there is some interest in the Xeon Phi among our collaborators, we
have created a basic HEP software development environment to
facilitate certain types of application and benchmark tests which can run
directly on the Phi card.

\subsection{Test Setup}
For the tests described here we used a Xeon Phi 7110P, with 61 in-order
cores (and up to 4-way Hyperthreading). The card was used with
a standard Intel Xeon box with 32 logical cores as the host.

\subsection{Software Environment}
There are two practical difficulties 
when setting up a software environment for the Phi. First, no compilation environment is available 
on the Phi itself, and cross-compilation must be done 
on the host. Second, currently only the Intel compiler
can be used for the compilation.

We chose to create what looks
like a normal SCRAM~\cite{SCRAMCHEP} environment for the user on the host system. A
user can create a SCRAM development area, with some number of previously
cross-compiled externals visible, and then (cross-)compile their own
code with SCRAM in a standard 
fashion. They need only switch to a shell on the Phi to actually run it.
We made a special CMSSW release which had cross-compiled externals
and just the subset of CMSSW packages which compile with the Intel
compiler.
%We have had a poor experience since many years in trying to use the
%Intel compiler for large C++ projects like those in HEP. 
%zlib bz2lib openssl expat readline sqlite db4 gdbm autotools python pcre xz libjpg xerces-c gccxml gsl ncurses pacparser photos pythia6 libpng libxml2 freetype lhapdf gmake cppunit fftw3 libuuid libtiff frontier\_client xrootd boost clhep hepmc root fastjet heppdt nspr vdt sigcpp pythia8 tauola geant4 charybdis herwig classlib elementtree roofit coral (not all 125 yet, though)
Most externals were built by just setting CXX=``icpc -fPIC -mmic'' and
CC=``icc -fPIC -mmic'' and --host=x86\_64-k1om-linux to configure scripts
for cross compilation. A few special cases include:
\begin{itemize}
\item Boost: Patched a couple of files and used TOOLSET intel
\item Python: Needed to be built twice, once for build system and once for Xeon Phi cross compilation.
\item Fastjet: Compiled without -msse3
\item GSL: Fixed configure script to not run test when cross-compiling
\item OpenSSL: Configured without -fstack-protector and --with-krb5-flavor
\item Root: Patched to build some executables without -mmic. Built without fftw3, castor and dcap dependency. Configured for linuxx8664k1omicc along with a couple of patches to use freetype and pcre from cms externals. Option -mmic passed to icc fortran compiler.
\end{itemize}

This work began using version 13.1.3 of the Intel compiler, however
it lacked sufficient C++11 support to compile CMSSW, and in
particular the core framework,
used by much of the rest of the code. We switched to version 14.0.0 when
it was released, and indeed the C++11 support had
improved. We were then stopped by another bug~\cite{ICCBUG}, but
369 out of 1106 CMSSW packages do
now compile. As the goal here was not necessarily to run full CMS 
applications, but rather to permit tests with more narrow benchmarks,
this is already an interesting software environment.

\subsection{Experimental Results}
\begin{figure}
\centering
\begin{subfigure}
  \centering
  \includegraphics[width=0.4\textwidth]{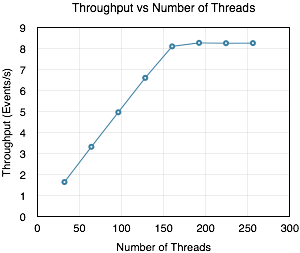}
\end{subfigure}
\begin{subfigure}
  \centering
  \includegraphics[width=0.4\textwidth]{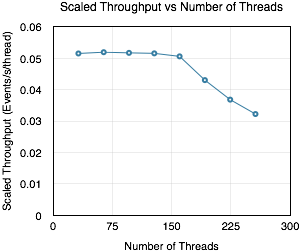}
\end{subfigure}
\caption{Toy event processing framework scaling tests on Xeon Phi}
\label{fig:toyfwk}
\end{figure}

Performance tests were run on the Xeon Phi using a toy version of the
threaded event processing framework being developed for CMS~\cite{CHEP13FWK}. The framework
uses Intel's Threading Building Blocks library to schedule the running of
both events and modules within an event concurrently. The toy framework
does not use actual CMS algorithms for the modules but instead emulates
the timing and dependencies between the algorithms based on measurements
from the full CMS framework. Each module in the toy
framework just does a simple numeric integration and continues the
integration for as long as the emulated CMS framework module would
run. Figure~\ref{fig:toyfwk} shows the toy framework scales linearly
until around 150 threads at which time it plateaus.  We believe we understand this behavior based on the Xeon Phi design. Although
the Xeon Phi has four hardware threads per core, the documentation states
that only two of the threads can run simultaneously. The third and fourth hardware threads are used in the cases where a running thread has a
memory access latency. In that case, the hardware can set aside that
waiting thread and use the core to process instructions for another
hardware thread. Therefore the plateau at 150 can be explained
by the fact that the numerical integral work does not require long
memory accesses and therefore we can saturate the hardware with less
than 3 hardware threads per core.

We also ported to the Xeon Phi a real world C++ application used
for the photon energy regression training in the search
for Higgs boson decaying to two photons. The application, based on ROOT and multithread-enabled via OpenMP, is very demanding in terms of CPU: when executed on a 4-way SandyBridge E5-4620, it fully uses the 64
cores (32 real plus 32 in HyperThreading), exceeding 6000\% CPU usage as
reported by {\it top}. The porting on the Phi was not straightforward
due to the inability to compile on the Phi: the initial ROOT
application was using a mixture of compiled libraries, and code compiled
on the fly by ROOT using ACLiC. While the libraries were cross-compiled on
the host system, the auto-compiled part could not be executed due to the
lack of a compilation environment on the Phi itself. We
had to revert to fully
compiled code, which meant losing some features of the ROOT
environment like fast turnaround code changes and 
debugging. The application was able to spawn
up to 243 threads on the Phi, but CPU occupancy, as reported by both
{\it top} and the {\it micsmc} tool, never exceeded 20\% of the
maximum theoretical load. Indeed the total testing time, found to be
about 6 hours on the Sandy Bridge, exceeded 24 hours, at which point
the process was killed by us; we estimate that it had reached by then
around 40\% of the application workflow. Further optimisations and
specific Phi changes are certainly possible, but the interesting test
for us was to check whether the porting of a working multi-threaded application, designed for a standard x86 architecture, would be just
a simple recompilation; from this point of view we cannot declare the test successful.

\section{General Tools Support}

\subsection{IgProf Profiler}
When comparing and optimizing for various architectures, understanding
the performance obtained in detail is as important as obtaining
overall benchmark numbers. For large C++
applications like those used in HEP, we have been using the
IgProf~\cite{IGPROF1,IGPROF2} profiling and analysis tool. We report
that as of version 5.9.10, IgProf itself has initial support for both 
performance sampling and memory profiling also on ARMv7 processors. 
There are currently still limitations from ARM stack unwinding as
implemented in libunwind, for speed in both modes, and due to
crashes in performance profiling mode. The next step will likely be
improvements to libunwind, much like those we contributed for similar problems on x86-64.

\subsection{Distributed MultiThreaded CheckPointing (DMTCP)}
In complex computing environments there are a number of use cases
for checkpointing the state of a running process to disk and restarting
it later. One technology providing this functionality is the
Distributed MultiThreaded CheckPointing (DMTCP) package~\cite{DMTCP}.
We have previously described the use cases which are interesting for HEP
and reported results~\cite{ACAT13DMTCP} for the use of checkpoint-restart
for HEP applications on x86-64 and on the Xeon Phi. From version 2.0.1,
we can now also report that DMTCP functions for HEP applications on ARMv7.

\section{Conclusions}
We report here on our evolving series of tests with the ARMv7 processor.
Single core performance is much lower for ARMv7 than x86-64, however the
performance per watt is much better for the ARMv7 processors. Here the
potential for use in scientific (general purpose) computing is clear.
As part of this work, we can also report successful ports of both
the IgProf profiler and the DMTCP checkpointing package to ARMv7.
We also report on our work to create a software development environment
to facilitate basic tests and benchmarking on the Xeon Phi. We now
have a reasonable software environment and report on a couple of initial
tests, however the potential for use in general HEP computing is not
yet clear.

\section*{Acknowledgments}
This work was partially supported by the National Science Foundation, under
Cooperative Agreement PHY-1120138 and Grant~OCI-0960978, and by the
U.S. Department of Energy.

\end{document}